\documentclass{article}

\PassOptionsToPackage{numbers, compress}{natbib}

\usepackage[preprint]{neurips_2022}

\usepackage[utf8]{inputenc} 
\usepackage[T1]{fontenc}    
\usepackage{hyperref}       
\usepackage{url}            
\usepackage{booktabs}       
\usepackage{amsfonts}       
\usepackage{nicefrac}       
\usepackage{microtype}      
\usepackage{xcolor}         
\usepackage{algorithm}
\usepackage{algpseudocode}
\usepackage{amsmath}
\usepackage{xfrac}
\usepackage{changepage}
\usepackage{amssymb}
\usepackage{todonotes}
\usepackage{multirow}
\usepackage{natbib}
\usepackage{amsthm}
\usepackage{enumitem,amsmath}
\usepackage{caption}
\usepackage{subcaption}
\theoremstyle{remark}
\newtheorem*{remark}{Remark}

\usepackage{kantlipsum}
\allowdisplaybreaks

\makeatletter
\newcommand\nocaption{%
    \renewcommand\p@subfigure{}
    \renewcommand\p@subtable{}
    \renewcommand\thesubfigure{\thefigure\alph{subfigure}}
    \renewcommand\thesubtable{\thetable\alph{subtable}}
}

\title{Pushing the limits of fairness impossibility: Who's the fairest of them all?}

\author{%
  Brian Hsu \\
  LinkedIn Corporation \\
  Sunnyvale, CA \\
  \texttt{bhsu@linkedin.com} \\
  \And
  Rahul Mazumder \thanks{Rahul Mazumder participated in this work within the scope of his consulting responsibilities at LinkedIn.} \\
  Massachusetts Institute of Technology \\
  Cambridge, MA \\
  \texttt{rahulmaz@mit.edu} \\
  \AND
  Preetam Nandy \\
  LinkedIn Corporation \\
  Sunnyvale, CA \\
  \texttt{pnandy@linkedin.com} \\
  \And
  Kinjal Basu \\
  LinkedIn Corporation \\
  Sunnyvale, CA \\
  \texttt{kbasu@linkedin.com} \\
}

\begin{document}
\maketitle

\begin{abstract}
  The impossibility theorem of fairness is a foundational result in the algorithmic fairness literature. It states that outside of special cases, one cannot exactly and simultaneously satisfy all three common and intuitive definitions of fairness - demographic parity, equalized odds, and predictive rate parity. This result has driven most works to focus on solutions for one or two of the metrics. Rather than follow suit, in this paper we present a framework that pushes the limits of the impossibility theorem in order to satisfy all three metrics to the best extent possible. We develop an integer-programming based approach that can yield a certifiably optimal post-processing method for simultaneously satisfying multiple fairness criteria under small violations. We show experiments demonstrating that our post-processor can improve fairness across the different definitions simultaneously with minimal model performance reduction. We also discuss applications of our framework for model selection and fairness explainability, thereby attempting to answer the question: {\emph{who's the fairest of them all?}}
\end{abstract}

\section{Introduction}
While fairness in machine learning has received significant attention in recent years, most existing works focus on one of the many fairness criteria \cite{barocas-hardt-narayanan}. Consequently, practitioners are perhaps left with no choice but to use their best judgment to apply a single fairness criterion. We suspect that the conflicting nature of existing mathematical definitions of fairness might have led to this undesirable practice of narrowing down fairness-related measurement and mitigation to one chosen definition. This is somewhat analogous to the trade-off between precision and recall while evaluating model performance \cite{buckland1994relationship}. Instead of choosing one of precision or recall to evaluate the performance of a classification model, practitioners often evaluate the trade-off and choose models that can maintain a certain level of precision while optimizing recall or vice-versa \cite{gordon1989recall}. In this paper, we provide a framework for explicitly addressing such trade-offs among multiple fairness criteria and model performance toward optimal model selection.

One of the most prolific examples of fairness in machine learning arose from the ProPublica recidivism study (\cite{angwin_larson_mattu_kirchner_2016}), in which a risk assessment tool called COMPAS was found to be biased against black defendants. But beyond the immediate implications in criminal justice, the study also prompted more general studies in algorithmic fairness and, in particular, led to a key result highlighted in \cite{Chouldechova17}, \cite{Hardtetal_NIPS2016}, and \cite{kleinbergFairnessTradeoff} which some colloquially refer to as the "impossibility theorem" in fairness \cite{MiconiImpossibility, KailashImpossibility}. This theorem essentially states that three common definitions of algorithmic fairness - demographic parity \cite{dwork2012fairness}, equalized odds \cite{Hardtetal_NIPS2016}, and predictive parity \cite{barocas-hardt-narayanan}, cannot be simultaneously satisfied outside of pathological situations. Several works following these initial results have therefore focused on satisfying one metric (\cite{Hardtetal_NIPS2016, NandyEOdds}) or proposing adaptable methods for different metrics (\cite{Celis2019, Rezaei_2020}). Yet, while we do not deny the conclusions of the impossibility theorem, we also believe there have not been sufficient efforts to reconcile the conflicting fairness definitions to the best extent possible. We fill this gap in the literature by translating the trade-offs among multiple fairness criteria and model performance into a constrained optimization problem and propose a post-processing methodology for simultaneously achieving approximate fairness in the conflicting definitions simultaneously.

We believe our framework would alleviate the practitioner from making the hard choice in choosing a particular metric. Instead, if they have a partial ordering of importance amongst the metrics (which many possess), our framework would explicitly allow them evaluate such trade-offs. The application which we highlight in Section \ref{sec:applications} discusses these in detail. 
Overall, we make three main contributions in this work:
\begin{enumerate}[leftmargin=*]
    \item We design a flexible optimization framework that returns a post-processing score transformation function that can make scores group-wise $\epsilon$-fair along three definitions (demographic parity, equalized odds, and predictive rate parity) simultaneously. This framework can be applied to any binary classifier that produces a continuous score, can be configured for singular or multiple metrics of fairness, and also can account for fairness vs. performance trade-offs in terms of ROC-AUC.
    \item We present a novel reformulation of this non-convex optimization problem as a Mixed Integer Linear Program (MILP)~\cite{wolsey1999integer}. This reformulation allows us to find provably globally optimal solutions. We further show that in practice, we can consistently find better solutions through our global optimization method compared to local optimization methods in a reasonable time.
    \item We discuss and extend our framework from a post-processing mechanism to a tool that can aid practitioners in better understanding their data and models' empirical fairness characteristics and trade-offs and compare these traits across models.
\end{enumerate}

The rest of the paper is organized as follows. In Section \ref{sec:mfopt} we mathematically define fairness metrics and the multiple fairness optimization problem. We discuss the optimal solution via MILP in Section \ref{sec:optimalSolution}. We try to answer the question of ``who's the fairest of them all?'' through our applications and experiments in Section \ref{sec:applications}. Finally, we conclude with a discussion in Section \ref{sec:conclusion}. We wrap up this section with a discussion of the related literature.

\textbf{Related Work:}
Early works \cite{Chouldechova17} exploring the conflicts between fairness definitions prove that for a binary predictor, predictive rate parity conflicts with equalized odds (\cite{Hardtetal_NIPS2016}) unless base rates are equal or the model is perfectly predictive. \citet{Chouldechova17} also considers trade-offs between the fairness definitions in the binary prediction case. \citet{kleinbergFairnessTradeoff} generalizes this study by showing that statistical (i.e., demographic) parity is also inconsistent with predictive rate parity and equalized odds. The same paper studies these inconsistencies in a more general bin-wise prediction setting and shows that approximate fairness definitions (predictive rate parity, equalized odds) can simultaneously hold but only under $\epsilon$-approximate equal base rates or $\epsilon$-approximate perfect performance. This work further proves that there is an inherent trade-off between fairness and loss.

Beyond impossibility theorem results, several works have focused on trade-offs between fairness and model performance \citep{Hardtetal_NIPS2016, Fish2016, Rezaei_2020, pmlr-v54-zafar17a}. They develop in-processing solutions aimed at reducing one metric while maintaining accuracy. A few authors have analyzed trade-offs or attempted to achieve multiple fairness. One example is \citet{pleiss2017fairness}, which shows that predictive rate parity and a relaxed form of equalized odds are reconcilable under a randomized prediction scheme. Another notable example is \citet{Celis2019}, which develops a flexible in-processing approach to achieve multiple types of fairness (potentially at the same time). Like \citet{Celis2019}, the framework we develop also aims for a flexible approach to focus on one or many fairness metrics simultaneously. However, our method is distinct in that it is a post-processing based solution and also more general as it works for continuous scores (rather than binary classification). Our work is most closely related to \citet{NandyEOdds} in terms of the underlying score transformation mechanism, and we leverage some of their methods. However, whereas \cite{NandyEOdds} only targets equalized odds, we go further and include demographic parity and predictive rate parity in our framework---this posits computational challenges, which we address by proposing novel methodology based on
integer programming.

As noted at the end of \cite{kleinbergFairnessTradeoff}, some open questions are how to optimally assign scores to satisfy multiple criteria when base rates are equal and additionally, how to satisfy predictive rate parity and either equal TPR or TNR when one cost outweighs the other. To our knowledge, no prior work has attempted to reconcile all three fairness conditions (demographic parity, equalized odds, and predictive rate parity) simultaneously with model performance through a post-processing framework. Although the post-processing methods tend to be less flexible for fairness-performance trade-offs than their in-processing counterparts, they can be much more easily added on top of any existing model training pipeline. This makes a post-processing approach modular, and in particular, more appealing in complex web-scale recommender systems that use (a combination of) certain prediction scores to rank a list of items.
\section{Multiple Fairness Optimization}
\label{sec:mfopt}

 We consider a binary classification problem where the $i$-th observation is characterized by their label $y_{i} \in \{0,1\}$, their group membership $g_{i} \in \mathcal{G}$, and model predicted probability (also known as a risk score \cite{kleinbergFairnessTradeoff}) $s_{i} \in [0,1]$ for $i = 1,\ldots,N$. The corresponding random variables are denoted as $Y$, $G$ and $S$ respectively. To set up the problem, we discretize the scores into nonempty bins $b \in \mathcal{B} := \{1,\ldots, |\mathcal{B}|\}$ by using, for example, a quantile transformation (we will denote $|\mathcal{B}|$ as $B$). Additionally, let $N^{[g]}_{b+}$ denote the number of group $g$ positive class ($y_{i}=1$) instances in bin $b$, $N^{[g]}_{b}$ denote the total number of instances of group $g$ instances in bin $b$, $N^{[g]}_{+}$ ($N^{[g]}_{-}$) be the total number of group $g$ positive (negative) instances, and $N^{[g]}$ be the total number of group $g$ instances. Lastly, our approach seeks to achieve fairness by moving instances from one bin to another bin: hence, we define variable $x^{[g]}_{bb'}$ as the probability of moving an instance of group attribute $g$ and score in bin $b$ into a new bin $b'$ \footnote{In applications, we can discretize the scores into $B$ bins with a quantile discretizer and consider how we can move them across bins. More bins allow for more granular interpretation of the transformed scores at the cost of us solving a harder problem and vice versa.}. In other words, for every group $g$, the collection $\{x^{[g]}_{bb'}\}_{b,b'}$ can be represented as a $B \times B$ transition matrix (with additional constraints, as discussed next).

For the optimization framework and the remainder of this paper, we translate a single fairness definition (e.g. equal true positive rate) as a constraint that controls for the worst-case violations across all bins. Below, we discuss different fairness constraints that we consider in our framework, and describe how they can be represented in terms of the optimization variables $\{x_{bb'}^{[g]}\}$.

\subsection{Fairness Constraints Under Binning Framework}\label{sec:constraints1}
\vspace{-.5em}
\textbf{Demographic Parity (DP):}
For simplicity, we assume that there are exactly two groups $g \in \{1,2\}$ as we formulate the fairness metrics, although everything discussed below can be applied for non-binary groups by using pairwise constraints, which we elaborate on in Appendix \ref{appendix: Non-binary groups}. Starting with demographic or statistical parity from \cite{dwork2012fairness}, this condition states that the model's predicted score is independent of group membership. This is equivalent to 
$$ P(S=s \mid G=1)=P(S=s \mid G=2).$$
Our version of this constraint uses bins $B$ to empirically approximate the probability $P(S=s \mid G = g)$ and we also relax the equality to an $\epsilon$-approximate equality (for some pre-specified $\epsilon > 0$). Therefore, after transforming the scores using $\{x^{[g]}_{bb'}\}_{b,b'}$ the $\epsilon_{DP}$-approximate DP can be expressed as the following as a linear constraint: 
\begin{equation}\label{eq:dp} \left|\frac{1}{N^{[1]}} \sum_{b\in \mathcal{B}} x^{[1]}_{bb'}N^{[1]}_{b} - \frac{1}{N^{[2]}} \sum_{b\in \mathcal{B}} x^{[2]}_{bb'}N^{[2]}_{b} \right| \leq \epsilon_{DP} \qquad \forall\;\; b' \in \mathcal{B},
\end{equation} 
where $N_b^{[g]}$ denote the number of observations from group $g$ in bin $b$ (before transformation), and $N^{[g]} = \sum_{b\in \mathcal{B}} N_b^{[g]}$. For reproducibility, $\epsilon_{DP}$ should be chosen to be larger than the approximation error $O(1/\sqrt{N^{[g]}})$ for replacing $P(S=s \mid G = g)$ with its empirical counterpart.  \\

\textbf{Equalized Odds (EOdds):}
The equalized odds condition for binary predictors, given in \citealt{Hardtetal_NIPS2016}, is a balance condition where the groups must have equal true positive and false positive rates. For continuous scores, it translates to having equal score distributions for each group conditional on their true labels \cite{NandyEOdds}:
$$P(S=s\mid Y=y, G=1)=P(S=s\mid Y=y, G=2) \qquad\text{for}~y\in\{0,1\}.$$
Like demographic parity, our empirical score bin version requires that the distribution of positive or negative instances be $\epsilon_{EOdds}$-approximately equal between groups in the new bins $b'$. Both equal true positive rate and false positive rate can be expressed as linear constraints, respectively:
\begin{equation}
\label{eq:eodds} 
\begin{aligned}
\left|\tfrac{1}{N^{[1]}_{+}} \sum_{b\in \mathcal{B}} x^{[1]}_{bb'}N^{[1]}_{b+} - \tfrac{1}{N^{[2]}_{+}} \sum_{b\in \mathcal{B}} x^{[2]}_{bb'}N^{[2]}_{b+} \right| &\leq \epsilon_{EOdds}~~\forall\;\; b' \in \mathcal{B}\\
\left|\tfrac{1}{N^{[1]}_{-}} \sum_{b\in \mathcal{B}} x^{[1]}_{bb'}N^{[1]}_{b-}- \tfrac{1}{N^{[2]}_{-}} \sum_{b\in \mathcal{B}} x^{[2]}_{bb'}N^{[2]}_{b-} \right| &\leq \epsilon_{EOdds}~~\forall\;\; b' \in \mathcal{B}.
\end{aligned}
\end{equation}

\textbf{Predictive Rate Parity (PRP):}
Lastly, we examine the predictive rate parity condition popularized in \cite{Chouldechova17}. This condition states that the probability of being a positive instance is independent of group membership when we condition on the score. Formally:
$$P(Y=1\mid S=s, G=1)=P(Y=1\mid S=s, G=2).$$
Using the empirical score bin framework, an approximate version of the above implies that the proportion of positive instances in each bin must be $\epsilon_{PRP}$-approximately equal among groups: 
\begin{equation}
\label{eq:prp} \left|\frac{\sum_{b\in \mathcal{B}} x^{[1]}_{bb'}N^{[1]}_{b+}}{\sum_{b\in \mathcal{B}} x^{[1]}_{bb'}N^{[1]}_{b}} - \frac{\sum_{b\in \mathcal{B}} x^{[2]}_{bb'}N^{[2]}_{b+}}{\sum_{b\in \mathcal{B}} x^{[2]}_{bb'}N^{[2]}_{b}}  \right| \leq \epsilon_{PRP}~~~\forall\;\; b' \in \mathcal{B}.
\end{equation}
Unlike constraints~\eqref{eq:dp} and~\eqref{eq:eodds}, which can be expressed as a linear function of the optimization variables $\{x^{[g]}_{bb'}\}$, condition~\eqref{eq:prp} yields bilinear terms and is in general a non-convex constraint. A main technical difficulty of our framework arises from this non-convex fairness constraint---Section \ref{sec:optimalSolution} presents an integer programming framework to handle this non-convexity, ensuring we can obtain a globally optimal solution to the resulting optimization problem.  
\begin{remark}
Our definition of fairness as the worst-case violation across all bins aims to resemble approximations of the respective probabilistic definitions but we have not found identical definitions in other works. We comment on the differences and discuss why it does not contradict the traditional impossibility theorem of \cite{kleinbergFairnessTradeoff} in Appendix \ref{appendix: comparison to other fairness}. 
\end{remark}

\subsection{MFOpt: Multiple Fairness Optimization Framework} \label{MFOpt}
We use constraints developed in Section~\ref{sec:constraints1} to state the multiple fairness optimization (MFOpt) problem:
\begin{subequations}\label{eq:master-problem}
\begin{align}
    & \underset{\{x^{[g]}_{bb'}\}_{b, b',g} }{\text{minimize}} 
    & & \sum_{g \in \mathcal{G}}\sum_{b\in \mathcal{B}}\sum_{b' \in \mathcal{B}} \left|\frac{N^{[g]}_{b}}{N}(\Bar{s}_{b}-\Bar{s}_{b'})x^{[g]}_{bb'} \right| \label{eq:objective} \\
    & \text{s.t.} 
    & & \sum_{b\in \mathcal{B}}x^{[g]}_{bb'}=1 \qquad \forall \;\; b' \in \mathcal{B},\; g \in \mathcal{G}  \label{eq:movementSum} \\
    &&& x^{[g]}_{bb} \geq 1-\xi \qquad \forall\;\; b \in  b' \in \mathcal{B},\; g \in \mathcal{G}  \label{eq:maxmovement} \\
    &&& x^{[g]}_{bb'} = 0 \qquad \forall\;\; b' \ s.t. \ |b' - b| \geq w,\; \forall g \in \mathcal{G} \label{eq:windowconst} \\
    &&& \text{Fairness Constraints:} \quad (\ref{eq:dp}), (\ref{eq:eodds}), (\ref{eq:prp}) \label{eq:allFairness} \\
    &&& \frac{\sum_{b\in \mathcal{B}} x^{[g]}_{bb'}N^{[g]}_{b+}}{\sum_{b\in \mathcal{B}} x^{[g]}_{bb'}N^{[g]}_{b}} \leq \frac{\sum_{b\in \mathcal{B}}  x^{[g]}_{b(b'+1)}N^{[g]}_{b+}}{\sum_{b\in \mathcal{B}} x^{[g]}_{b(b'+1)}N^{[g]}_{b}} \qquad \forall\;\; b' \in \{ 1,...,B-1\},~g \in \mathcal{G} \label{eq:rankOrder} \\
    &&& 0 \leq x^{[g]}_{bb'} \leq 1,~~\forall\;\; b,b', g. \label{eq14}
    \end{align}
\end{subequations}
Above, the optimization variables are the group-specific movement probability $x^{[g]}_{bb'}$ terms and all remaining terms are problem data and/or configurable hyperparameters. Starting with the objective (\ref{eq:objective}), we define $\Bar{s}_{b}$ as the midpoint score in the bin and hence the objective is the product of the movement distance $\Bar{s}_{b}-\Bar{s}_{b'}$ weighed by the fraction of total samples moved $\sfrac{N^{[g]}_{b}}{N}$ and the amount of movement $x^{[g]}_{bb'}$. (\ref{eq:movementSum})~states that the total movement out of bin $b$, including the movement back to itself, must sum up to $1$ and along with (\ref{eq14}) ensures that $x^{[g]}_{bb'}$ represent probabilities in a transition matrix. (\ref{eq:maxmovement}) states that the total movement from bin $b$ back to itself must be lower bounded by hyperparameter $\xi$. This parameter controls how far we allow the new scores to stray from the original and is necessary to prevent zero denominators in \ref{eq:prp} and \ref{eq:rankOrder}. Constraint (\ref{eq:windowconst}) represents window constraints to restrict extreme movements of scores beyond $w$ bins. Constraint (\ref{eq:allFairness}) are the fairness constraints in Section~\ref{sec:constraints1}. (\ref{eq:rankOrder}) ensures that we preserve the rank-ordering of the scores and expected values, which is desirable for comparing bins against each other. While constraints (\ref{eq:windowconst}), (\ref{eq:rankOrder}) are not fairness-related, we add them to retain the utility of the solution. The predictive rate parity constraint (\ref{eq:prp}) and (\ref{eq:rankOrder}) both introduce non-convexities into Problem~\eqref{eq:master-problem}. These two constraints also require us to assume overlap, or that each bin contains at least one member from each group. Without overlap, predictive rate parity is undefined since it is not possible to compare expectations across groups for a given bin and demographic parity is likely violated as it means one group has 0 probability of landing in a given bin. 

We close this section with two major benefits of this framework compared to inprocessing solutions such as adding fairness regularization (\cite{pmlr-v54-zafar17a,Fish2016,Rezaei_2020}) or using an entirely different fairness-based model (\cite{Celis2019}, \cite{ZhangAdvDebiasing}). First, the number of optimization variables scales in the order of $\mathcal{O}(|\mathcal{G}||\mathcal{B}|^2)$ while the number of constraints scale in the order of $\mathcal{O}(|\mathcal{B}||\mathcal{G}|^2)$. This is significant as it entails that our framework can be applied in arbitrarily large data settings as long as score-based binning is possible. Applying the transformation is equally tractable, as it only requires binning observations and making independent draws from a multinomial distribution with $B$ possible outcomes.

The second benefit of our framework is that it returns a highly interpretable solution as it returns one optimized $B\times B$ transition matrix per group. Hence given a newly scored instance, several facts can be read from the corresponding row of the matrix such as the likelihood of moving to a specific bin $b$, moving into any higher or lower bin, etc. These probabilities can also be controlled via constraints as shown with the window constraints (\ref{eq:windowconst}) and max movement constraints (\ref{eq:maxmovement}). This interpretability is an advantage over model regularization frameworks, where it is difficult to know how an individual's score might change when switching from the base model to a fair model. 
\section{Finding Optimal Solutions via Mixed Integer Programming (MIP)} \label{sec:optimalSolution}
The primary difficulty of the above optimization problem are the predictive rate parity constraint (\ref{eq:prp}) and rank-order constraint (\ref{eq:rankOrder}) which turn the problem non-convex. Non-convex constrained optimization is generally NP-hard and traditional methods that seek locally optimal solutions include gradient-based interior point optimization (\cite{Wchter2006OnTI}), sequential quadratic programming (\cite{Gill2012SequentialQP}), or algorithms specific to quadratically-constrainted-quadratic-programs (QCQPs) such as operator splitting methods, semidefinite relaxations, etc. (see \cite{ParkQCQP} for an overview). 

Rather than pursuing a locally optimal solution, we propose a novel reformulation of the problem into a tractable mixed-integer-linear-program (MILP) which can be solved to global optimality (\cite{CastroNMD},  \cite{CASTROpiecewise}). Our reformulation grants two benefits over traditional locally optimal solvers. First, global strategies theoretically enable us to find the best possible solution. Second, as our problem primarily scales with the number of bins, it is practically tractable when utilizing the power of modern MIP solvers. 

\subsection{Tractable reformulations for computational efficiency}

We first observe that a direct reformulation of the fractional terms into bilinear terms in constraints (\ref{eq:prp}) and (\ref{eq:rankOrder}) will lead to bilinear terms in the order of $\mathcal{O}(B^3)$. We show that we can reduce this to $\mathcal{O}(B)$ bilinear terms through a substitution that exploits the problem structure. Next, we take advantage of the vastly reduced number of bilinear terms to apply the normalized multiparametric disaggregation technique (NMDT, \cite{Andrade2019}) which we explain in Section \ref{sec: fractionalLP}. This allows us to approximate products of continuous variables as products of binary variables, which can be easily linearized and handled by MIP solvers. Importantly, this transformation of  $xy$ (i.e, product) terms requires upper and lower bounds for $x$ and $y$ and we propose a method in Section~\ref{sec: fractionalLP} for generating and tightening these bounds by solving fractional linear program subproblems (\cite{CharnesCooper}). Taken together, the reduction of bilinear terms combined with the bound-tightening procedure enable us to effectively apply the NMDT methodology and transform the problem from a non-convex QCQP to an MILP that can be solved to global optimality.

\subsubsection{Step 1: Reducing number of bilinear variables} 
We reduce the number of bilinear variables in our problem by making a substitution for the fraction term by introducing new optimization variables $v_{b}^{[g]} \geq 0$ (to represent a sum) and $t_{b}^{[g]} \geq 0$ (to represent the fractional quantity) as additional variables (see Appendix \ref{appendix: nmdt} for more details). We can then use them to write equivalent constraints with only $\mathcal{O}(B)$ bilinear terms. Let,
\begin{align*}
    v^{[g]}_{b'} = \sum\nolimits_{b \in \mathcal{B}} x^{[g]}_{bb'}N^{[g]}_{b} \qquad\text{and}\qquad  t^{[g]}_{b'}v^{[g]}_{b'} = \sum\nolimits_{b \in \mathcal{B}} x^{[g]}_{bb'}N^{[g]}_{b+}  \qquad \forall\;\; b' \in \mathcal{B}. 
\end{align*}
Then we have the following:
\begin{align*}
    \text{Constraint}~(\ref{eq:prp}) &\Longleftrightarrow \left| t^{[1]}_{b'} - t^{[2]}_{b'} \right| \leq \epsilon_{PRP} \;\; \forall\;\; b' \in \mathcal{B},\\ \text{Constraint}~(\ref{eq:rankOrder}) &\Longleftrightarrow t^{[g]}_{b'} \leq t^{[g]}_{b'+1} \;\; \forall\;\; b' \in \{1, \ldots, B-1\}
\end{align*}

\subsubsection{Step 2: NMDT and bound tightening through fractional LP subproblems} \label{sec: fractionalLP}
\noindent {\bf Linearizing bilinear terms (NMDT):} We show how we can model each bilinear term 
$t_{b}^{[g]}v_{b}^{[g]}$ by using a binary expansion for the continuous variables $t_{b}^{[g]},v_{b}^{[g]}$, and by observing that the product of binary variables can be modeled via 
integer programming (see~\cite{LiChangPolynomial,TELES2013613} for reference). To this end, we make use of the NMDT transformation~\cite{Andrade2019}. 
We recap this method below as formulated in \cite{Andrade2019}\footnote{The open-source implementation can be found in \url{https://github.com/joaquimg/QuadraticToBinary.jl} (MIT License) which we utilize.} Given any bounded optimization variable $x\in [x_{L}, x_{U}]$, and precision factor $p$, a negative integer, we can represent this variable exactly as $x = (x_{U} - x_{L}) \lambda + x_{L}$ where
$$\lambda = \sum_{l\in \{-p, \ldots, -1\}} 2^{l}z_{l} + \Delta\lambda$$
where $0 \leq \Delta\lambda \leq 2^{p}$ is a remainder term and $z_{l} \in \{0,1\}$ are binary optimization variables. Dropping the remainder term $\Delta\lambda$ gives us the approximate form and product forms of $xy$ become dot products of several integer variables, which can be effectively handled via modern MIP solvers, such as \cite{gurobi}. Although we are solving an approximation (e.g. precision of $1e^{-4}$) this is not a practical problem since it is precise enough for reasonable choices of $\epsilon$ and we do not expect the constraints to hold exactly when we apply the post-processor on the testing data.

\noindent {\bf Bounds on $v_{b}^{[g]},t_{b}^{[g]}$  via Fractional LPs:} A key requirement to apply NMDT is that all optimization variables in the bilinear terms ($t^{[g]}_{b}$ and $v^{[g]}_{b}$ in our case) must be bounded and we need to be able to accurately estimate these bounds (i.e., tighter bounds leads to faster runtimes \cite{Andrade2019}). We propose obtaining these lower and upper by minimizing and maximizing respectively, the sub-problems while keeping all fairness constraints except the quadratic constraints. Firstly, note that bounds on $v^{[g]}_{b}$ can be solved as a simple LP (details in Appendix \ref{appendix: nmdt}). Meanwhile, obtaining a bound on $t^{[g]}_{b}$ requires solving a nonlinear problem due to fact that $t^{[g]}_{b}$ represents a fractional objective (ratio of two affine terms of optimization variables). However, we observe that we can apply the Charnes-Cooper transformation~\cite{CharnesCooper} to reformulate the nonlinear problem into a simple LP (details in Appendix \ref{appendix: nmdt}).

\subsection{Choice of algorithm: QCQP (heuristic) vs MIP (optimal solution)}
In this section we show the results and benefits of our reformulation from a non-convex QCQP to an MILP. In Table \ref{solutionComparisonTable}, we take each dataset, create a 60/40 train-test split, train a grid-searched random forest model, and score the training data. Next, we discretize the scores into bins, parameterize the problem (\# bins, $\epsilon$, max movement, window size, solve time) on the scored training data, and compare our MILP solution solved by Gurobi (\cite{gurobi}) against the QCQP problem solved by IPOPT (\cite{coinor}), which is a generic interior-point log-barrier penalty method for nonlinear constrained optimization. We discuss the datasets and problem parameters for all experiments in the Appendix \ref{appendix: experiments data}. For each metric such as $AUC$, we use $AUC_{INT}$ and $AUC_{IP}$ to denote the average result of applying the interior-point (INT, for short) or integer programming (IP) method, respectively. The metrics used are the objective value, optimality gap ($\% \Delta$), \footnote{The optimality gap is defined $\% \Delta = \frac{Upper Bound - Lower Bound}{Upper Bound}$ where the upper bound is the best feasible solution and the lower bound is produced by the branch-and-bound method. The INT method does not have the benefit of providing lower bounds, hence we use the bound produced by the IP method to compute this.} and $AUC$\footnote{We describe how we computed the AUC for the bin-wise probabilities in Appendix \ref{appendix: auc}.}. We also report the statistical significance of the improvement based the $p$-value from the Wilcoxon signed-rank test to determine if $\% \Delta_{IP} \leq \% \Delta_{INT}$ is a consistent result. Bold figures indicate statistical significance w.r.t. 1 standard deviation\footnote{We are limited in the number of trials we can run and actively chose to prioritize the variety of datasets we apply on rather than a large number of trials for a single dataset. As such, we expect relatively large standard errors but we reflect the consistency of our method through the p-value.} \footnote{See Appendix \ref{appendix: experiments data} for an explanation of the stark underperformance of IPOPT on the ACS Coverage and COMPAS datasets.}.
\vspace{-0.5cm}

\begin{center}
\begin{table}[!ht]
  \caption{Interior Point Solution vs. MIP Solution}
  \label{solutionComparisonTable}
  \centering
  \resizebox{\textwidth}{!}{\begin{tabular}{lllllllll}
    \toprule
    Dataset & $Obj_{INT}$ & $Obj_{IP}$ & $\%\Delta_{INT}$  & $\%\Delta_{IP}$ & $p$-value & $AUC_{INT}$ & $AUC_{IP}$ \\
    \midrule
    ACS Income &  2.0809 &  1.9682 &   15.076 ± 6.461 &  10.621 ± 3.402 &  \textbf{0.0029} &  0.9041 &  0.9044 \\
    ACS Insurance &  0.9769 &  0.9599 &    3.432 ± 0.225 &   \textbf{1.715 ± 0.169} &  \textbf{0.0010} &  0.7411 &  0.7413 \\
    ACS Mobility &  2.4580 &  2.3781 &     5.37 ± 0.803 &   \textbf{2.193 ± 0.138} &  \textbf{0.0010} &  0.7971 &  0.7973 \\
    ACS Poverty &  2.0693 &  2.0526 &    3.756 ± 0.435 &   \textbf{2.972 ± 0.324} &  \textbf{0.0010} &  0.8440 &  0.8440 \\
    ACS Coverage  &  8.9361 &  1.9665 &   79.711 ± 0.782 &   \textbf{7.878 ± 2.207} & \textbf{0.0010} &  0.5420 &  0.8149  \\
    ACS Travel &  2.3935 &  2.3859 &    2.554 ± 0.254 &    2.242 ± 0.28 &  \textbf{0.0010} &  0.7725 &  0.7725 \\
    Heart Disease &  1.8871 &  1.3035 &  26.385 ± 17.401 &    \textbf{3.81 ± 0.864} &  \textbf{0.0010} &  0.8302 &  0.8629 \\
    COMPAS &  7.4551 &  3.1300 &   62.88 ± 13.407 &  \textbf{17.055 ± 7.482} &  \textbf{0.0010} &  0.5143 &  0.7378 \\
    \bottomrule
  \end{tabular}}
\end{table}
\end{center}

\vspace{-0.5cm}
As the results show, the MIP reformulation consistently beats the interior point solver applied on the raw optimization problem, even when solving for only 10 minutes. We also observe that in most cases, regardless of the method we choose, we can quickly find near optimal solutions that are high performing in the AUC sense. This is a significant result as it means that even a locally optimal solution to our optimization problem can yield a practically useful post-processing result. Additional experiments on these datasets showing the effectiveness of MFOpt with respect to performance and fairness in a training/testing data scenario can be found in Appendix \ref{appendix: Additional Experiments}.

We conclude this section by reiterating two benefits of the reformulation. First, solving a MIP method yields lower bounds that can be used to prove optimality or otherwise gauge the quality of a feasible solution. Second, by framing the problem as a MIP, we can always theoretically continue improving the solution to optimality based on the acceptable time limits. There are other applicable global optimization methods, such as spatial branch and bound. We considered these solutions but ultimately opted for a MIP approach due to the maturity and availability of solvers (see Appendix \ref{appendix: commentary on other solutions} for details).

\vspace{-2.5mm}
\section{Applications}
\label{sec:applications}
\begin{figure}[ht!]
\caption{\label{fig1:frontierImg}Efficient frontier of solutions for ACS West Insurance data}
\centering
\includegraphics[width=\textwidth]{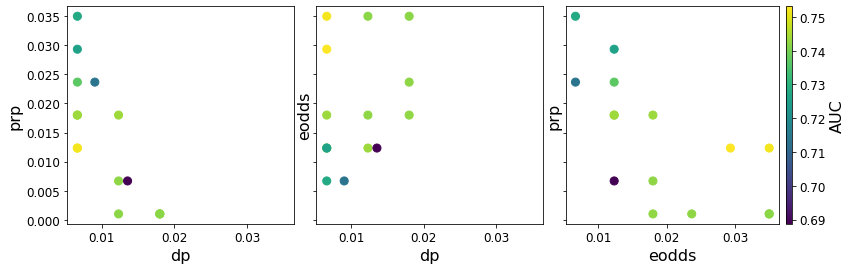} 
\end{figure}

In this section, we illustrate a few methods of applying our framework and how it can be used to help model developers select and understand models from a fairness perspective. To apply these procedures, we first require developing an efficient frontier of fairness solutions to understand which $\epsilon$ configurations are feasible for a given model type and dataset. To generate this frontier, we solve the problem over a grid of parameters $\epsilon_{DP}, \epsilon_{EOdds}, \epsilon_{PRP}$. Each feasible solution will yield a point $s \in \{(AUC, \epsilon_{DP}, \epsilon_{EOdds}, \epsilon_{PRP})\}$ and the collection of non-dominated points from the solution set yields a efficient frontier. We show the 2-d profile shots of our 4-d fairness surface in Figure \ref{fig1:frontierImg} as an example, where the axes represent one of three fairness metrics and the point color gradient represents AUC. In theory, we could obtain a true Pareto-optimal frontier since we have devised a method of obtaining globally optimal solutions. However, we generate this frontier using IPOPT due to practical limitations as we are solving a $7 \times 7 \times 7$ grid of $\epsilon$ parameters.

\begin{table}[h!]
\vspace{-0.5cm}
\centering
         \caption{Performance Fairness Trade-off Analysis}
  \label{tradeoffAnalysis}
\resizebox{0.6\textwidth}{!}{\begin{tabular}{ccccccc}
    \toprule
    Trade... & For... & $s^{*}_{AUC}$ & $s^{*}_{\epsilon_{DP}}$ & $s^{*}_{\epsilon_{EOdds}}$ & $s^{*}_{\epsilon_{PRP}}$ \\
    \midrule
    Base & Base & 0.7434 & 0.0123 & 0.018 & 0.0123 \\ 
    AUC & $\epsilon_{DP}$ & - & - & - & - \\
    AUC & $\epsilon_{Eodds}$ & - & - & - & - \\
    AUC & $\epsilon_{PRP}$ & 0.7422 &  0.0123 &  0.0180 &  0.0067 \\
    $\epsilon_{DP}$ & $\epsilon_{PRP}$ & 0.7436 &  0.0152 &  0.0123 &  0.0067 \\
    $\epsilon_{Eodds}$ & $\epsilon_{PRP}$ & - & - & - & - \\
    $\epsilon_{Eodds}$ & $\epsilon_{DP}$ & 0.7436 &  0.0152 &  0.0123 &  0.006 \\
    \bottomrule
  \end{tabular}}
\end{table}

\subsection{Understanding fairness tradeoffs}
After the Pareto frontier is generated, the modeler can pick an operating point $s$ based on the desired AUC and tolerable fairness violations $\epsilon$. However, when communicating fairness properties to stakeholders and accounting for potential changes in strategy, it can be helpful to additionally understand the cost of further increasing fairness in terms of $\epsilon_{DP}, \epsilon_{EOdds}, \epsilon_{PRP}$. 

This tradeoff can be understood by looking at the characteristics of points on the frontier near the operating point. Suppose we are at an operating point $s$. If we want to trade $AUC$ for $\epsilon_{EOdds}$, then we would find a point $s'$ with at least as good $\epsilon_{DP}, \epsilon_{PRP}$ but worse $AUC$ and better $\epsilon_{EOdds}$. More generally, if we pick trade performance/fairness characteristic $c$ (cost) for characteristic $b$ (benefit), then we hold all other factors constant and find a point with better $b$ and worse $c$. We illustrate this in the Table \ref{tradeoffAnalysis}. The first row shows a hypothetical operating point while the following rows show other points on the efficient frontier that we could move to when we make a certain trade. Blank rows indicate that no such point was found that permits the desired trade-off. This trade-off perspective can enable developers to better understand and communicate the costs to performance or other fairness metrics when trying to close the disparity in one fairness metric.

\subsection{Performance Comparison} 

Lastly, we compare our framework against two methods and show that we can satisfy fairness constraint(s) just as well or better, while obtaining significantly stronger performance. Our comparison is done as follows (details in \ref{appendix: experiments data}), in each iteration we randomly split and pre-process the data, tune and fit a random forest model, and score the training and testing data to get the base scores $\hat{y}_{0}$. Next, we run the methodology that we are comparing against (i.e. build a model or apply the postprocessor) to get method scores $\hat{y}_{m}$. We then bin the outputs of the base model and compared method and compute the AUC along with the fairness metrics ($\epsilon_{0}, \epsilon_{m}$). Next, we solve our constrained optimization problem on the training data, where we set the parameters $\epsilon$ to $\frac{1}{2}min(\epsilon_{0}, \epsilon_{m})$. After optimizing, we can apply the optimal solutions $x^{[g]}_{bb'}$ to assign new bins in the testing data based on the original score bins. One method of assigning a group $g$ instance with score $s \in b$ (denote as $s_{b}^{[g]}$) would be to randomly draw from a multinomial distribution parameterized by probabilities $(x^{[g]}_{b1}, x^{[g]}_{b2}, \ldots, x^{[g]}_{b|\mathcal{B}|})$. We propose alternative methods to this stochastic assignment method in the Appendix~\ref{appendix:bin to score}. Lastly, we compute the resulting AUC and fairness metrics on remapped bins for the testing data (and do the same for $\hat{y}_{0}$ and $\hat{y}_{m}$ on the testing data). These figures are shown in Table \ref{tbl:methodComparison}. We only show the results on the test set due to space constraints and have placed the results for the training set in Appendix \ref{appendix: experiments data}. Bold figures indicate that a metric is statistically significant to 1 standard deviation. 

First, we compare our framework against the in-processing framework in \citet{Rezaei_2020}, which is a robust optimization-based logistic regression model for reducing equality of opportunity violation. We found that this method works well compared to standard logistic regression and managed to decrease fairness violations while maintaining the similar performance. It improves in $\epsilon_{Eodds}$ compared to a random forest as well. However, its weakness is that the underlying model is still logistic regression and therefore has limited expressiveness. In comparison, MFOpt can be applied on top of any model class and thus the performance advantages of more flexible models are better maintained.  

Next, we compare our framework against the post-processing framework in \citet{pleiss2017fairness}. In this method, the authors use randomization to maintain the model's calibration while simultaneously satisfying a relaxed equalized odds condition (whereby a linear combination of TPR and TNR are satisfied). Again, we see that this method maintains close performance as the base model, successfully shrinks equalized odds violations, and even decreases demographic parity violations too. However, it results in large violations of predictive rate parity based on our definition in contrast to our method.

\begin{table}
\centering
         \caption{Comparison with other fairness methods}
         \label{tbl:methodComparison}
\begin{tabular}{cccccc}
    \toprule
    \multirow{2}{*}{Method} & \multirow{2}{*}{Metric} & 
      \multicolumn{3}{c}{Testing Data} \\
    & & Base & Method & MF-Opt \\ 
    \midrule
    \multirow{4}{*}{Rezaei}
    & $AUC$ &   0.7471 ± 0.003 &  0.6619 ± 0.0022 &    0.747 ± 0.003 \\ 
    & $\epsilon_{DP}$ &  0.0117 ± 0.0014 &  0.0124 ± 0.0013 &   \textbf{0.0088 ± 0.001} \\ 
    & $\epsilon_{EOdds}$ &   0.0266 ± 0.007 &  0.0291 ± 0.0059 &  \textbf{0.0167 ± 0.0029} \\
    & $\epsilon_{PRP}$ &   0.109 ± 0.0145 &  0.1091 ± 0.0143 &  0.0986 ± 0.0133 \\
    \addlinespace[0.2cm]
    \multirow{4}{*}{Pleiss} 
    & $AUC$&  0.8319 ± 0.0033 &  0.8149 ± 0.0087 &   0.831 ± 0.0032 \\ 
    & $\epsilon_{DP}$ &  0.0212 ± 0.0016 &  0.0137 ± 0.0016 &  \textbf{0.0106 ± 0.0011} \\ 
    & $\epsilon_{EOdds}$ &  0.0329 ± 0.0042 &   0.023 ± 0.0038 &  \textbf{0.0142 ± 0.0028} \\
    & $\epsilon_{PRP}$ &  0.1465 ± 0.0178 &  0.4147 ± 0.1537 &  0.1547 ± 0.0293 \\
 \bottomrule
  \end{tabular}
\end{table}

\subsection{Who's the fairest of them all?}

In industry, machine learning model selection is guided by many factors including performance, speed, interpretability, among others. Yet, the fairness dimension is commonly overlooked unless the developers specifically induce it in their model. Even then, picking a specific fairness metric to optimize for can be a nebulous task. Rather than focus on a single metric, we propose a simple and intuitive method of gauging a model's efficiency in trading between different fairness definitions.

To do so, we first construct the frontier for the two models in question and then filter all points on the efficient frontier with tolerable performance $AUC \geq AUC_{min}$. Next, we find the point on the respective frontiers with minimum Euclidean distance to the origin. The model with the shorter distance to their frontier can then be declared as the model that has better tradeoff properties. We do not show an example due to lack of space, but remark that this procedure and the trade-off analysis can be useful when a developer is iterating between models that were not designed for fairness, but wants a model that can be flexibly made more fair through the postprocessing framework that we propose. As fairness requirements may change over time, the model that can yield the best tradeoffs between different definitions can offer the most overall utility. 

\section{Discussion}
\label{sec:conclusion}

In our study, we have devised a flexible, tractable, and interpretable post-processing method which we apply to push the limits of the impossibility theorem. We show that while theoretical limitations remain undisputed, there is a path forward to practically reconciling the conflicting fairness definitions. These results extend the findings of \cite{Rodolfa_2021}, which state that the trade-off of fairness and accuracy are negligible in practice. Our work reinforces this claim but also adds on that trade-offs between fairness definitions can be negligible as well. For further research, one area is to improve the consistency of the PRP violation reduction, as we observed the largest standard error in reducing this metric. This could be due to us using random forests for all experiments, which is known to be an uncalibrated model \cite{RFUncalibrated}. One method to address this is therefore first calibrating the model through other methodologies such as Platt scaling (\cite{Platt99probabilisticoutputs}) or binning-based calibration (\cite{Kumar2019}) before applying MFOpt. We could also consider incorporating uncertainty in the training data through principled approaches such as stochastic or robust optimization.

\bibliographystyle{abbrvnat}
\bibliography{Fairness}

\clearpage
\appendix

\section{Mapping from bins to scores} \label{appendix: Non-binary groups}
Extending the MFOpt framework to non-binary groups requires creating pairwise constraints for all groups in each bin to achieve the same fairness effect. This means that the number of constraints scaled in the order of $\mathcal{O}(|\mathcal{B}||\mathcal{G}|^2)$ where in practice, we expect $|G|$ to be in the range of 3-4. This pairwise procedure does not affect the scaling of the number of optimization variables, which remains at $\mathcal{O}(|\mathcal{B}|^{2}|\mathcal{G}|)$. However, we note that the scaling factor of the constraints is not a major bottleneck due to our problem reformulation as presented in Section \ref{sec:optimalSolution}. Without the reformulation, the pairwise constraints would generate $\mathcal{O}(|\mathcal{B}|^{3}|\mathcal{G}|^2)$ bilinear terms from the cross-multiplication of the numerator/denominator sums and return a much more computationally intensive problem.

\section{Mapping from bins to scores} \label{appendix:bin to score}
As mentioned, the most straightforward method of applying the score transformation after solving the optimization problem is to sample from a multinomial distribution. However, this is a less granular approach as we are assuming that all observations in the bin are indistinguishable. To overcome this, we recommend the idea proposed in (\cite{NandyEOdds}) which is a linear projection. This strategy proposes that if an observation with score $s$ falling into a bin $a$ with upper and lower bounds $[a_{l}, a_{u}]$ gets mapped from the random draw into a new bin $b_{1}$ with bounds $[b_{1l}, b_{1u}]$, then we assign it a linearly interpolated score given by:
$$ s'=b_{1l}+\frac{s-a_{l}}{a_{u}-a_{l}}(b_{1u}-b_{1l}) $$
This allows us to maintain rank-ordering of scores that receive the same assignment from $a$ to $b$. 

A more deterministic manner of mapping from bin to score would be to take the expected score mapping. After solving the optimization problem, we know the transitions probabilities $a$ to \{$b_1$, $b_2$, \ldots, $b_B$\} (denoted as $P(a\longrightarrow b_{i})$) based on the optimization variables and from the previous method, we also know the score assignment if $a$ were moved into $b_{i}$ (denote as $s_{i}$). Hence, a deterministic map would transform score $s$ to $s'=\sum_{i\in B}s_{i}P(a\longrightarrow b_{i})$.

\section{Computing AUC from bins and using AUC as an objective} \label{appendix: auc}
Another idea we leverage from (\cite{NandyEOdds}) is the Riemann approximation of AUC from the bins. Essentially, ROC AUC be approximated by the FPR at bin $k$ and TPR of the cumulative bins $b \in \{k,\ldots, B\}$. Another consideration is that we could have changed the objective to maximizing AUC rather than minimizing score movement. However, in our experience, maximizing AUC (quadratic objective) as the objective led to a harder time finding better feasible solutions compared to minimizing score movement (linear objective). 

\section{Details on the fractional LP subproblem for bound tightening} \label{appendix: nmdt}
We elaborate on the methodology in Section \ref{sec: fractionalLP}. Recall that our goal is to find bounds for:

\begin{align*}
    v^{[g]}_{b'} = \sum_{b \in \mathcal{B}} x^{[g]}_{bb'}N^{[g]}_{b} \qquad\text{and}\qquad  t^{[g]}_{b'}v^{[g]}_{b'} = \sum_{b \in \mathcal{B}} x^{[g]}_{bb'}N^{[g]}_{b+}  \qquad \forall\;\; b' \in \mathcal{B}. 
\end{align*}

Where $t^{[g]}_{b'}$ is meant to represent the fractional quantity:
$$ t^{[g]}_{b'} = \frac{\sum_{b\in \mathcal{B}} x^{[g]}_{bb'}N^{[g]}_{b+}}{\sum_{b\in \mathcal{B}} x^{[g]}_{bb'}N^{[g]}_{b}} = \frac{\sum_{b\in \mathcal{B}} x^{[g]}_{bb'}N^{[g]}_{b+}}{v^{[g]}_{b'}} $$
We will do this by fixing $\bar{g}$ and $\bar{b}$ such that we first tighten bounds for $v_{\bar{b}}^{\bar{g}}$ and then use the optimal solution to tighten bounds for $t_{\bar{b}}^{\bar{g}}$. First, it is easy to see that maximizing/minimizing  $v_{\bar{b}}^{\bar{g}}$ is an LP as we have dropped the quadratic constraints, leaving us with a linear objective and linear constraint set. Now let $v^{\bar{g}*}_{\bar{b},min/max}$ represent the optimal values of the min/max objective for $v_{\bar{b}}^{\bar{g}}$. We now turn to bounding $t_{\bar{b}}^{\bar{g}}$ which has the same linear constraints but a fractional (nonlinear) objective. To deal with this, we utilize the Charnes-Cooper transformation (\cite{CharnesCooper}). Essentially, this reformulation trick handles the denominator by removing it from the objective and passing it to all constraints while maintaining linearity. To illustrate this in detail, we first define new optimization variables:
\begin{equation}
    \xi^{[g]}_{bb'} = \frac{x^{[g]}_{bb'}}{\sum_{b \in \mathcal{B}} x^{[\bar{g}]}_{b\bar{b}}N^{[\bar{g}]}_{b}} \qquad \phi_{\bar{b}}^{[\bar{g}]} = \frac{1}{\sum_{b \in \mathcal{B}} x^{[\bar{g}]}_{b\bar{b}}N^{[\bar{g}]}_{b}} \label{eq18}
\end{equation}
 Using (\ref{eq18}), we can express the min/max problem for $t^{[\bar{g}]}_{\bar{b}}$ as problem (\ref{problem:FLP}).
 
 \begin{align}
\label{problem:FLP}
\allowdisplaybreaks
    & \underset{\xi^{[g]}_{bb'}, \phi}{\text{Min or Max}}
    & & t^{[\bar{g}]}_{\bar{b}} = \sum_{b \in \mathcal{B}} N^{[\bar{g}]}_{b+}\xi^{[\bar{g}]}_{b\bar{b}} \nonumber  \\ 
    & \text{subject to} 
    & & \sum_{b \in \mathcal{B}} \xi^{[g]}_{b\bar{b}}N^{[\bar{g}]}_{b}=1 \nonumber \\
    &&& \xi^{[g]}_{bb'} \geq (1-m)\phi \qquad \forall\;\; b=b'\nonumber \\
    &&& \xi_{bb'} = 0 \qquad \forall\;\; b' \ s.t. \ |b' - b| \geq w \nonumber\\
    &&& \left|\frac{1}{N^{[1]}} \sum_{b \in \mathcal{B}} \xi^{[1]}_{bb'}N^{[1]}_{b} - \frac{1}{N^{2}} \sum_{b \in \mathcal{B}} \xi^{[2]}_{bb'}N^{[2]}_{b} \right| \leq \epsilon_{DP}\phi \qquad \forall\;\; b' \in B \\
    &&& \left|\frac{1}{N^{[1]}_{+}} \sum_{b \in \mathcal{B}} \xi^{[1]}_{bb'}N^{[1]}_{b+} - \frac{1}{N^{[2]}_{+}} \sum_{b \in \mathcal{B}} \xi^{[2]}_{bb'}N^{[2]}_{b+} \right| \leq \epsilon_{EOdds}\phi \qquad \forall\;\; b' \in B \nonumber \\
    &&& \left|\frac{1}{N^{[1]}_{-}} \sum_{b \in \mathcal{B}} \xi^{[1]}_{bb'}N^{[1]}_{b-} - \frac{1}{N^{[2]}_{-}} \sum_{b \in \mathcal{B}} \xi^{[2]}_{bb'}N^{[2]}_{b-} \right|\nonumber
    \leq \epsilon_{EOdds}\phi \quad \forall\;\; b' \in B \nonumber \\
    &&& \frac{1}{v^{g*}_{b'max}} \leq \phi \leq \frac{1}{v^{g*}_{b'min}} \qquad 0 \leq \xi^{[g]}_{bb'} \leq \frac{1}{v^{g*}_{b'min}} \nonumber
\end{align}

 By solving these subproblems and taking the objective value as bounds for $t^{[g]}_{b}$, we can reduce the feasible region of the problem and enhance our solutions. The sub-problems are bounded and if any of them are infeasible, then it also implies that the MFOpt problem is also infeasible as we drop the PRP constraints in these sub-problems:

\section{Experiment data descriptions and problem parameters} \label{appendix: experiments data}
We use three primary data sources for our experiments, the more recently developed American Community Survey (ACS) data as well as two more classical datasets, Heart Disease and COMPAS. We elaborate on each dataset in this section. The ACS data is a dataset made publicly available by the US Census Bureau. Specifically, Ding et. al \cite{DingRetiringAdult} have created an excellent Python package\footnote{https://github.com/zykls/folktables (MIT License)} that enables users to pull model-ready data (for a requested year and geographic region) for a set of pre-defined binary classification tasks, such as predicting high income, health insurance coverage, whether they move or not, among others. The tasks are detailed in the paper and we use all of the pre-defined tasks without any additional modification except for Employment. We do not use the Employment task because of the assumption detailed in Section \ref{MFOpt} regarding overlap. Experiments with this task occasionally yielded models that did not have overlap which made this task unsuitable for demonstrating our methodology. We reiterate that this is not a practical issue if one just ignored the non-overlapping bins, but requires a lengthy and technical fairness interpretation that we felt were beyond the purpose of our study. In terms of time and geography, we use 2020 data for all experiments while the geography varies. In the experiments shown on Table \ref{solutionComparisonTable}, we use the West Coast US states (California, Oregon, Washington). In \ref{tbl:methodComparison} we wanted a larger dataset as we required a sufficiently large testing split, hence we used the West Coast States ('CA', 'OR', 'WA', 'NV', 'AZ') with the "ACS Mobility" dataset and a 60/40 train-test split for the inprocessing comparison and East Coast States ('ME','NH','MA','RI', 'CT','NY', 'NJ', 'DE','MD','VA', 'NC','SC','GA','FL') and same with the "ACS Poverty" dataset for the postprocessing comparison. There was no particular reason for selecting these geographies aside from obtaining a large enough sample that we can feasibly run multiple trials on. Though we are using census data there is no PII information nor any endangerment to the subjects in the data. However, we note that in practice, it is important to exercise caution and equity in picking groups to mitigate for, as selective mitigation of favored groups by a malicious practitioner can result in underperformance for deserving groups. 

The Heart Disease Dataset (\cite{heartDiseaseData}) is a publicly available dataset where the task is to predict whether or not an individual has heart disease. Most applications of this data use the standard processed "Cleveland" data and we use sex as the group variable. We could not find a standard and preprocessed version of this data and did it ourselves by one-hot-encoding categorical variables. 

The COMPAS dataset is based on the recidivism study noted in (\cite{angwin_larson_mattu_kirchner_2016}). We use the preprocessed version made available in the publicly available AIF360 package (\cite{aif360-oct-2018}) \footnote{https://github.com/Trusted-AI/AIF360} without any additional modification. In this dataset, the task is to predict whether or not an individual will recidivate and we use ethnicity as the group variable.

We list the problem parameters used to create the results in Table \ref{solutionComparisonTable} in Table \ref{problemParameterTable}. We use the same parameters for all tasks. A particular note is that in Table \ref{solutionComparisonTable}, the IPOPT method performs very poorly compared on the COMPAS and ACS Coverage data. This is because across the 10 trials we ran, the IPOPT algorithm had frequently failed to converge within the 10-minute time limit for these two datasets. From the outputs, we saw that convergence failure is accompanied by heavy violation of the predictive rate parity constraint (demographic parity and equalized odds are still satisfied) and a high objective value. The exact reason for frequent failure in these two datasets is unclear, however, we hypothesize that it is due to a relatively high predictive rate parity gap in the data which led to numeric issues. Such failures were not observed in the IP formulation while both solvers were provided the exact same problem parameters.

All experiments were run on a MacBook Pro with a 2.4GHz 8-Core Intel Core i9 processor with 32 GB RAM. We did not use the GPU for solving. Data, preprocessing steps, and the random forest models utilize Python's scikit-learn (\cite{scikit-learn}, BDS License) package. The optimization model is coded through Julia's JuMP package (\cite{DunningHuchetteLubin2017} MPL License). We use the Gurobi (\cite{gurobi} Academic License) and IPOPT (\cite{coinor} Eclipse Public License) solvers for all problems.

Due to lack of space, we only showed the method comparison Table \ref{tbl:methodComparison} for the testing data. We show the results on the training data in Table \ref{tbl:methodComparisonTrain}.
\begin{center}
\begin{table}
  \caption{Experiment Problem Parameters}
  \label{problemParameterTable}
  \centering
  \begin{tabular}{ccccccc}
    \toprule
    \# Trials & Bins & $\epsilon$ & Max Movement & Window Size & Solve Time & Precision \\
    \midrule
    10 & 50 & 0.03 & 0.5 & 13 & 600s & 1e-5\\
    \bottomrule
  \end{tabular}
\end{table}
\end{center}

\begin{center}
\begin{table}
  \caption{Comparison with other fairness methods}
  \label{tbl:methodComparisonTrain}
  \centering
  \begin{tabular}{cccccc}
    \toprule
    \multirow{2}{*}{Method} & \multirow{2}{*}{Metric} & 
      \multicolumn{3}{c}{Train} \\
    & & Base & Method & MF-Opt \\ 
    \midrule
    \multirow{4}{*}{Rezaei} 
    & $AUC$ &   0.7471 ± 0.003 &  0.6619 ± 0.0022 &    0.747 ± 0.003 \\ 
    & $\epsilon_{DP}$ &  0.0117 ± 0.0014 &  0.0124 ± 0.0013 &   0.0088 ± 0.001 \\ 
    & $\epsilon_{EOdds}$ &   0.0266 ± 0.007 &  0.0291 ± 0.0059 &  0.0167 ± 0.0029 \\
    & $\epsilon_{PRP}$ &   0.109 ± 0.0145 &  0.1091 ± 0.0143 &  0.0986 ± 0.0133 \\
    \addlinespace[0.2cm]
    \multirow{4}{*}{Pleiss} 
    & $AUC$&  0.8314 ± 0.0045 &  0.8145 ± 0.0104 &  0.8306 ± 0.0044 \\ 
    & $\epsilon_{DP}$ &  0.0208 ± 0.0029 &  0.0145 ± 0.0023 &  0.0105 ± 0.0008 \\ 
    & $\epsilon_{EOdds}$ &  0.0325 ± 0.0062 &   0.0257 ± 0.005 &  0.0144 ± 0.0017 \\
    & $\epsilon_{PRP}$ &  0.1405 ± 0.0214 &  0.4149 ± 0.1824 &  0.1319 ± 0.0204 \\
 \bottomrule
  \end{tabular}
\end{table}
\end{center}

\section{Comparison to other fairness definitions} \label{appendix: comparison to other fairness}
We compare our bin-wise worst-case fairness definition with other fairness definitions seen in literature and explain why it does not contradict previous impossibility theorem results. First, since we are considering score bins, our definition is a generalization of the definitions in (\cite{Chouldechova17}), (\cite{Hardtetal_NIPS2016}) and (\cite{Celis2019}), which consider fairness metrics for binary $\{0, 1\}$ classifiers or assume that there is a threshold for mapping probabilities to {0,1} outcomes. In these cases, the overall FPR/FNR can be computed and EOdds refers to the equality of those rates. Our framework is a generalization since the same fairness metrics for binary classification can be achieved by specifying that all scores be moved into exactly one of two bins (representing \{0,1\} predictions) under our framework. 

Since we are dealing with binned scores, our fairness definitions more resemble those seen in \cite{kleinbergFairnessTradeoff}, which also has a notion of binned "risk assignments". The critical difference in fairness definitions is that Kleinberg's paper utilize the sum of scores in each bin compared against the number of positive or negative instances. Under this scheme, predictive rate parity refers to having the sum of scores be equal to the number of positive instances and true positive rate refers to the expected score of the positive instances in each bin (and analogously with FPR). The key difference is that rather than the sum of scores, our definitions are based on the expected number of \{0,1\} instances moved into each bin, irrespective of the instance's original scores. As such, we are not faced with the same strict fairness trade-offs. 

\section{Commentary on other solution methods and solvers} \label{appendix: commentary on other solutions}
While investigating a solution to our nonconvex problem, we considered another global integer programming based approach known as spatial branch and branch (SBB), which relies on a combination of spatial partitioning and solving local partitions using McCormick relaxations (\cite{Mccormick1976}, \cite{CASTROpiecewise}) and other outer approximation variants (\cite{BurerMIQCP}). In our testing, Gurobi's nonconvex QCQP solver, which applies these SBB heuristics, worked remarkably well despite being relatively new and was sometimes able to beat both the interior point solution and the MIP solution. Open-source solver SCIP (\cite{BestuzhevaEtal2021OO}) also features a gender nonconvex SBB solver that works reasonably well. However, our main goal was to provide a widely accessible method of solving the problem to global optimality and as of writing, there are significantly more developed open-source MILP solvers, such as SCIP, HiGHS (\cite{huangfuHiGHs}), and CBC (\cite{coinor}), than SBB solvers. Another reason we opted for the MILP approach is that we saw more potential in the NMDT reformulation for taking advantage of our reformulation and bound tightening procedure. Nonetheless, our bound tightening procedure is theoretically beneficial for both methods and as other open-source algorithms/solvers for SBB become more developed, such as Couenne (\cite{coinor}) and Alpine (\cite{NagarajanAlpine}), we encourage a future re-evaluation of solution methods and comparisons.

Finally, we note that we chose Gurobi in our experiments for its speed and effectiveness since we are repeatedly solving many problems. We acknowledge that Gurobi is a very powerful commercial solver and the results solved over 10 minutes may be worse with open source solvers such as HiGHS (\cite{huangfuHiGHs}). Nonetheless, the important fact is that all MIP solvers target the global optimum and hence even less powerful solvers can yield strong solutions given more time.

\section{Additional Experiments Using Expected Assignment on Testing Data} \label{appendix: Additional Experiments}
We list additional experiments focusing on the performance of our method on the testing data in Tables \ref{tbl: addExp Travel} to \ref{tbl: addExp Public}. All results are based on 20 trials that are run with a similar procedure in the comparison section \ref{sec:applications} except that we compute the testing metrics based on expected assignment rather than stochastic assignment (explained below), which we feel better reflects the \textit{average} performance of MFOpt. In each trial, we tune a random forest via grid-search, find the base fairness violations, set up the parameters of MFOpt to reduce the violations by a half, and optimize. Then, we compute the results (AUC and fairness violations) on the testing data (baseline with no modifications vs. MFOpt) and show the 1-Standard deviation error margins as well as the $p$-value corresponds to a one-sided Wilcoxon signed rank test which evaluates if the distribution of differences of the Base - MFOpt stats (higher AUC, lower fairness violations) is symmetric around zero (null) or instead favors the base (alternative). 

The major difference in this evaluation compared to the results shown in section \ref{sec:applications} is that in section \ref{sec:applications}, we assign observations to bins in the testing data based on random draws from a multinomial distribution (explained in Appendix \ref{appendix:bin to score}). However, a single draw per train-test split may not properly reflect the expected performance of MFOpt, even if we average over 20 trials. Instead, we feel a more accurate representation of the expected performance is if we apply the expected bin assignment to obtain the post-movement number of $\{0,1\}$ and total samples for each bin in the testing set. Under this procedure, we do not move individual observations across bins, but rather move all of them together. Concretely, suppose that we have optimized parameters $(\hat{x}^{[g]}_{1b'}, \hat{x}^{[g]}_{2b'}, \ldots, \hat{x}^{[g]}_{Bb'})$, which represented the probabilities of observations from each bin moving into bin $b'$. Then we propose that the number of \{1\} outcomes at bin $b'$ after the expected bin assignment procedure is:

$$\hat{N}^{[g]}_{b'+} = \sum_{b\in \mathcal{B}} \hat{x}^{[g]}_{bb'} N^{[g]}_{b+}$$

The same computation applies for the expected number of negative samples $\hat{N}^{[g]}_{b'-}$ and total samples $\hat{N}^{[g]}_{b'}$, which we use to compute the AUC and fairness violations and report in the tables. We find that across different datasets, the decrease in AUC is miniscule in terms of both absolute amount and variance (less than 1\%). We obviously do not expect better AUC from the MFOpt solution compared to the unconstrained model and thus this result is remarkable as it indicates that some degree of fairness can be afforded practically for free under our framework. The second observation is that we can reduce all three fairness metrics simultaneously and consistently across all datasets, as we find $p$-values below 0.05 in all cases. We do observe variance of the PRP violation is relatively higher than that of DP or EOdds. We noted this in our conclusion Section \ref{sec:conclusion} as an area for future work and provide some hypotheses for methods that can address this inconsistency. Nonetheless, improving conflicting definitions of fairness simultaneously is another significant result as it provides empirical evidence that there is a path forward towards multiple fairness. 

\vspace{-0.5cm}
\begin{center}
\begin{table}[hbt!]
  \caption{ACS West Travel} \label{tbl: addExp Travel}
  \centering
  \begin{tabular}{cccc}
    \toprule
    Metric & Base & MFOpt & Wilcoxon $p$-value\\
    \midrule
    AUC &  0.7439 ± 0.0039 &  0.7437 ± 0.0039 &  0.999999 \\
    DP &  0.0313 ± 0.0057 &  0.0208 ± 0.0045 &  0.000001  \\
    EOdds &  0.0404 ± 0.0055 &  0.0268 ± 0.0081 &  0.000001 \\
    PRP &  0.1743 ± 0.0326 &  0.1481 ± 0.0306 &  0.000001 \\
    \bottomrule
  \end{tabular}
\end{table}
\end{center}
\vspace{-1cm}

\begin{center}
\begin{table}[hbt!]
  \caption{ACS West Income}
  \centering
  \begin{tabular}{cccc}
    \toprule
    Metric & Base & MFOpt & Wilcoxon $p$-value\\
    \midrule
    AUC &  0.8907 ± 0.0012 &  0.8902 ± 0.0016 &  1.000000\\
    DP &  0.0172 ± 0.0021 &  0.0126 ± 0.0019 &  0.000001  \\
    EOdds &   0.0362 ± 0.006 &  0.0231 ± 0.0035 &  0.000001 \\
    PRP &  0.1994 ± 0.0745 &  0.1507 ± 0.0295 &  0.000024 \\
    \bottomrule
  \end{tabular}
\end{table}
\end{center}
\vspace{-1cm}

\begin{center}
\begin{table}[hbt!]
  \caption{ACS West Mobility}
  \centering
  \begin{tabular}{cccc}
    \toprule
    Metric & Base & MFOpt & Wilcoxon $p$-value\\
    \midrule
    AUC &  0.7413 ± 0.0033 &  0.7412 ± 0.0033 &  0.999284 \\
    DP &  0.0183 ± 0.0057 &  0.0149 ± 0.0033 &  0.008591  \\
    EOdds &  0.0469 ± 0.0153 &  0.0327 ± 0.0068 &  0.000031 \\
    PRP &   0.197 ± 0.0318 &  0.1738 ± 0.0311 &  0.000024\\
    \bottomrule
  \end{tabular}
\end{table}
\end{center}
\vspace{-1cm}

\begin{center}
\begin{table}[hbt!]
  \caption{ACS West Insurance}
  \centering
  \begin{tabular}{cccc}
    \toprule
    Metric & Base & MFOpt & Wilcoxon $p$-value\\
    \midrule
    AUC &  0.7183 ± 0.0028 &  0.7182 ± 0.0029 &  0.999916\\
    DP &  0.0603 ± 0.0089 &  0.0322 ± 0.0041 &  0.000001  \\
    EOdds  &  0.0676 ± 0.0072 &  0.0579 ± 0.0062 &  0.000052 \\
    PRP &    0.3732 ± 0.11 &   0.205 ± 0.0446 &  0.000018 \\
    \bottomrule
  \end{tabular}
\end{table}
\end{center}
\vspace{-1cm}

\begin{center}
\begin{table}[hbt!]
  \caption{ACS West Poverty}
  \centering
  \begin{tabular}{cccc}
    \toprule
    Metric & Base & MFOpt & Wilcoxon $p$-value\\
    \midrule
    AUC &  0.8319 ± 0.0039 &  0.8316 ± 0.0039 &  1.000000\\
    DP &  0.0246 ± 0.0038 &  0.0154 ± 0.0017 &  0.000001\\
    EOdds  &  0.0396 ± 0.0086 &  0.0226 ± 0.0024 &  0.000001\\
    PRP &  0.1305 ± 0.0172 &   0.1173 ± 0.019 &  0.000018 \\
    \bottomrule
  \end{tabular}
\end{table}
\end{center}
\vspace{-1cm}

\begin{center}
\begin{table}[hbt!]
  \caption{ACS West Public Coverage}\label{tbl: addExp Public}
  \centering
  \begin{tabular}{cccc}
    \toprule
    Metric & Base & MFOpt & Wilcoxon $p$-value\\
    \midrule
    AUC &  0.7932 ± 0.0016 &  0.7924 ± 0.0018 &  1.000000\\
    DP &    0.03 ± 0.0041 &  0.0204 ± 0.0028 &  0.000001 \\
    EOdds  &  0.0403 ± 0.0061 &  0.0236 ± 0.0028 &  0.000001\\
    PRP &   0.159 ± 0.0245 &  0.1443 ± 0.0297 &  0.029129\\
    \bottomrule
  \end{tabular}
\end{table}
\end{center}

\section{Role of Bins Parameter and Ablation Study} 
\label{appendix: Bin ablation study}

In this section, we first comment on the choice of the number of bins, $|\mathcal{B}|$, as a hyperparameter and then show results of additional testing with respect to the choice of the bin parameter. In practice, the choice of $|\mathcal{B}|$ should reflect the required bin granularity of the outputs for usage in, for example, rank-ordering or threshold-based classification. Our work therefore approaches choosing $|\mathcal{B}|$ purely from a computational complexity perspective as the number of optimization variables scales in the order of $\mathcal{O}(|\mathcal{G}||\mathcal{B}|^{2})$. We used $|\mathcal{B}|=50$ across our experiments as we found that this parameter gave a reasonable degree of granularity in the resulting bins while also being computationally tractable enough to solve (our MIP solver could frequently find solutions with <10\% optimality gap within 10 minutes). From a more theoretical perspective, the choice of $|\mathcal{B}|$ determines how well we can estimate the score transformation function, with higher $|\mathcal{B}|$ giving us better estimates in the training data. However, having too few samples within each bin (in the training data) results in low bias but high variance estimates of the score transformation function and the corresponding performance/fairness metrics, leading to bad generalization on the testing set. Having too few bins can, on the other hand, lead to an under-parameterized score transformation function. While it is clear that $|\mathcal{B}|$ should scale with respect to the per-bin sample size, we consider a rigorous analysis of this choice to be out of scope as a current limitation of our work. We hence recommend choosing $|\mathcal{B}|$ in practice based on the desired granularity of results and/or treating it as a parameter to optimize via cross-validation.

To run an experiment to study the effects of changing $|\mathcal{B}|$, we must first define $|\mathcal{B}|$-agnostic metrics to evaluate the post-processing output in terms of performance and fairness. The primary difficulty is that since fairness is defined on a bin-wise basis, evaluating the worst-case violation for example using $|\mathcal{B}|=10$ is not comparable to $|\mathcal{B}|=50$. Hence, we do the following: First, after optimizing for $x^{[g]*}_{bb}$, we apply the linear interpolation based mapping as described in Appendix \ref{appendix:bin to score} so that we have new scores $s'$. To compare the accuracy across different $|\mathcal{B}|$, we compute the ROC and precision-recall (PR) AUC based on $s'$. To compare the fairness, we discretize $s'$ in 100 bins (regardless of which $|\mathcal{B}|$ we used) and compute the worst case violation for each fainess metric across all 100 bins. The reason we use 100 bins to assess fairness is that we specifically use a large dataset (500k+ instances in both training and testing) to obtain more granular metrics. To generate the results, we use the ACS Income data as presented in previous sections. The difference is that we now use data across all states to obtain a large sample for a single random 60/40 train-test split, each with over 500k samples. Following the procedure from before, we train and tune a random forest model, score the training and testing data, and proceed with our evaluation methodology. Since the linear interpolation mapping methodology utilizes stochastic draws, we sample 300 draws per bin and compute the 1 standard-deviation errors shown in the tables.

Table \ref{tbl: bin ablation performance} shows the performance metrics across different $|\mathcal{B}|$, where we show the average metric and 1-standard deviation error margins. We observe that performance is very similar across the different choices, but is maximized at the higher bin counts of of 50-60. The performance also appears to plateau beyond a certain point, suggesting that it is not necessary to select a large  $|\mathcal{B}|$ for performance purposes. Interestingly, we also see that the solutions are near-optimal based on the optimality gap. This shows that although the number of variables scales with $|\mathcal{B}|$, it does not necessarily preclude us from high quality solutions. On the contrary, the optimality gap for lower bin ranges such as 25 is larger with lower performance. This could be due to the fact that since we are optimizing over the same $\epsilon$ fairness criteria, having fewer degrees of freedom with smaller bins makes it more difficult to find feasible solutions. 

Table \ref{tbl: bin ablation fairness} shows the results for fairness, where we show the average worst fairness violation and the 1-standard deviation error margins. The story is less clear from this angle but we emphasize that training the problem for different $|\mathcal{B}|\neq100$ bins using the same $\epsilon$ parameters and evaluating it on $|\mathcal{B}|=100$ for fairness is an unintended method of using our framework which we are only doing to have comparable results for the ablation study. Here, we see that the average metrics are similar for different parameters with $|\mathcal{B}|=40$ having the best overall result. Demographic parity violation is surprisingly minimized at $|\mathcal{B}|=30$ in this example, but given the other fairness and performance metrics $|\mathcal{B}|=40$ appears to be the best choice. Under this method of evaluation, the fairness parameters do not appear to be very sensitive to the choice of $|\mathcal{B}|$.

\begin{center}
\begin{table}[hbt!]
  \caption{Bin Ablation Study (Performance) - ACS Income} \label{tbl: bin ablation performance}
  \centering
  \begin{tabular}{cccc}
    \toprule
    Num. Bins & Optimality Gap & ROC AUC & PR AUC \\
    \midrule
    25 &   0.1483 &  0.8712 ± 0.0001 &  0.7933 ± 0.0002\\
    30 &   0.0672 &     0.8733 ± 0.0 &  0.7969 ± 0.0002\\
    35 &   0.0326 &     0.8739 ± 0.0 &  0.7981 ± 0.0002\\
    40 &   0.0095 &      0.874 ± 0.0 &  0.8001 ± 0.0001\\
    45 &   0.0096 &     0.8742 ± 0.0 &     0.8 ± 0.0001\\
    50 &   0.0144 &     0.8743 ± 0.0 &  0.8002 ± 0.0001\\
    55 &   0.0150 &     0.8742 ± 0.0 &  0.8005 ± 0.0001\\
    60 &   0.0212 &     0.8743 ± 0.0 &  0.8011 ± 0.0001\\
    \bottomrule
  \end{tabular}
\end{table}
\end{center}

\begin{center}
\begin{table}[hbt!]
  \caption{Bin Ablation Study (Fairness) - ACS Income} \label{tbl: bin ablation fairness}
  \centering
  \begin{tabular}{cccc}
    \toprule
    Num Bins & Demographic Parity & Equalized Odds & Predictive Rate Parity \\
    \midrule
    25 &  0.0174 ± 0.0004 &  0.0224 ± 0.0008 &   0.1447 ± 0.013\\
    30 &  0.0144 ± 0.0004 &  0.0206 ± 0.0005 &  0.1702 ± 0.0101\\
    35 &  0.0177 ± 0.0003 &  0.0235 ± 0.0019 &  0.1535 ± 0.0149\\
    40 &  0.0179 ± 0.0004 &  0.0195 ± 0.0005 &   0.1295 ± 0.013\\
    45 &  0.0174 ± 0.0003 &  0.0234 ± 0.0008 &  0.1406 ± 0.0182\\
    50 &  0.0215 ± 0.0003 &  0.0239 ± 0.0004 &  0.1456 ± 0.0143\\
    55 &  0.0165 ± 0.0004 &   0.021 ± 0.0006 &  0.1323 ± 0.0127\\
    60 &  0.0179 ± 0.0004 &  0.0237 ± 0.0007 &  0.1503 ± 0.0178\\
    \bottomrule
  \end{tabular}
\end{table}
\end{center}

\end{document}